# $O_xH_{2x+1}^+$ Clusters: A New Series of Non-Metallic Superalkali Cations by Trapping $H_3O^+$ into Water

Ambrish Kumar Srivastava

*Department of Physics, DDU Gorakhpur University, Gorakhpur-273009, Uttar Pradesh, India*

E-mail: ambrishphysics@gmail.com; aks.ddugu@gmail.com

## Abstract

The term 'superalkali' refers to the clusters with lower ionization energy than alkali atoms. Typical superalkali cations include a central electronegative core with excess metal ligands, $OLi_3^+$, for instance, which mimic the properties of alkali metal ions. We report a new series of non-metallic superalkali cations, $O_xH_{2x+1}^+$ ($x$ = 1-5) using *ab initio* MP2/6-311++G(d,p) level. These cations are designed by successive replacement of H-ligands of hydronium cation ($OH_3^+$) by ammonium ($OH_3$) moieties followed by their geometry optimization. The resulting $O_xH_{2x+1}^+$ clusters, which can be expressed in the form of $OH_3^+\cdots(x\text{-}1)H_2O$ complexes, possess a number of electrostatic as well as partially covalent H-bonds, with the interacting energy in the range 5.2-29.3 kcal/mol as revealed by quantum theory of atoms in molecules analyses. These cations are found to be stable against deprotonation as well as dehydration pathways, and their stability increases with the increase in $x$. Interestingly, the vertical electron affinities ($EA_v$) of $O_xH_{2x+1}^+$ clusters decreases rapidly from 5.16 eV for $x$ = 1 to 2.67 eV for $x$ = 5, which suggest their superalkali nature. It is also possible to continue this series of non-metallic superalkali cations for $x > 5$ with even lower $EA_v$, down to an approximated limit of 1.85 eV, which is obtained for $OH_3^+$ trapped into water cavity implicitly using polarizable continuum model. The findings of this study will not only provide new insights into structure and interactions of $O_xH_{2x+1}^+$ clusters but also reveal their novel properties, which can be exploited their interesting applications.

**Keywords:** Superalkali, Hydronium, Water solvent, Vertical electron affinity, Hydrogen bonds, *Ab initio* calculations.

## 1. Introduction

Superalkali cations, possessing lower ionization (IE) energy than alkali atoms, with a generic formula $XLi_{k+1}^+$ such as $FLi_2^+$, $OLi_3^+$, $NLi_4^+$ etc. were proposed by Gutsev and Boldyrev [1]. Apart from their experimental verification by Wu *et al.* [2], such species have been explored continuously by theoretical calculations to date [3-13]. Owing to low IE, they possess strong reducing properties and can be employed in the design of a variety of charge transfer species with unusual properties. The application of superalkalis in the design of supersalts [14-17], superbases [18-20], alkalides [21-23], etc. have already been reported. The single-electron reduction of $CO_2$ using superalkalis has been extensively studied [11, 24, 25]. Sun *et al.* [10] suggested the design of a new series of superalkali cations, $XM_2^+$ by taking typical superalkalis (M= $FLi_2$, $OLi_3$, and $NLi_4$) as ligands. Most of the superalkalis reported, thus far, contain metal atoms, such as Li, as ligands [1-7]. Hou *et al.* [8] made an attempt to study non-metallic superalkalis, exploring $F_2H_3^+$, $O_2H_5^+$, $N_2H_7^+$, etc. Recently, Giri *et al.* [9] have explored the superalkali behavior of $P_7R_4^+$ cations using $P_7^{3-}$ Zintl core decorated with organic ligands [R = Me, $CH_2Me$, $CH(Me)_2$ and $C(Me)_3$]. The rational design of superalkalis presented by Zhao *et al.* [11] explored another non-metallic superalkali cation, namely $C_5NH_6^+$. Despite these efforts, the reports on the superalkali nature of non-metallic cations are relatively scarce. Apart from the unusual structures they possess, these non-metallic cations are also interesting due to the fact that they have tendency to mimic the property of alkali metal cations. In this study, we propose a series of non-metallic superalkali cations systematically using *ab initio* approach.

$OH_3^+$, hydronium ion, is popularly known as Eigen cation [26]. It is an abundant molecular ion in the interstellar medium, found in diffuse [27] as well as dense [28] molecular clouds, which is produced by protonation of water ($H_2O$). Similarly, $O_2H_5^+$ is commonly referred to as Zundel cation [29, 30]. The nature of $H^+$ in aqueous media has been widely discussed in literature [31-33]. There exist many theoretical as well as experimental studies on $H^+(H_2O)_n$ clusters for various values of $n$, which are recently reviewed by Fournier *et al.* [34]. Most of these studies were mainly focussed on the structures and spectroscopic features of $H^+(H_2O)_n$ clusters. Lee and co-workers suggested [35-38] that the Eigen motif appears in $n$ = 1, 3–8 while the Zundel motif is found in $n$ = 2, 6–8. There is no systematic study on the electronic properties of these clusters to the best of our knowledge. In particular, the superatomic aspects of these clusters are unrevealed to date. Note that $OH_3^+$ is a non-metallic analog of $OLi_3^+$, a typical superalkali cation [1, 2]. Therefore, it is interesting to observe whether it possesses superalkali behavior. More interestingly, we investigate the superalkali behavior of $O_xH_{2x+1}^+$ cations ($x$ = 1-5), which are generally obtained by replacing an H-atom by $OH_3$ moiety successively.

## 2. Computational details

The equilibrium structure $O_xH_{2x+1}^+$ cations and their isomers were obtained by the geometry optimization at second order Møller-Plesset perturbation theory (MP2) [39] and 6-311++G(d,p) basis set without any symmetry constraints in the potential energy surface. The total energies of $O_xH_{2x+1}^+$ cations were obtained by single-point calculations at the same level. In order to elucidate the superalkali nature, we have computed vertical electron affinity ($EA_v$) of $O_xH_{2x+1}^+$ cations by the difference of the total electronic energy of equilibrium structure of cations and single-point energy of respective neutral species at cationic geometry. This provides the energy required to attach an electron to $O_xH_{2x+1}^+$ cations, which is equivalent to the energy released to extract an electron from $O_xH_{2x+1}$, i.e., the IE of $O_xH_{2x+1}$ species. The effect of water solvent on the $OH_3^+$ has been analyzed implicitly using polarizable continuum model (PCM) of Tomasi and coworkers [40]. All *ab initio* computations were carried out using Gaussian 09 suite of programs [41]. The quantum theory of atoms in molecules (QTAIM) analyses [42] were performed using AIMAll software package [43].

## 3. Results and discussion

### 3.1. Structural features

$OH_3^+$ is an isoelectronic to ammonia ($NH_3$), possessing trigonal pyramidal $C_{3v}$ symmetry with O-H bond lengths of 0.978 Å and bond angle of 112.1°. These are in good agreement with corresponding values of 0.974 Å and 113.6° obtained by infrared spectroscopy [44]. In $OH_3^+$, central O has a negative charge of -0.755 |e| and all H atoms have +0.585 |e| each. We replace one of H in $OH_3^+$ by $OH_3$, which leads to the structure of $O_2H_5^+$ as displayed in Fig. 1. This Zundel cation possesses $C_2$ symmetry and appears as $H_2O$-$H^+$-$OH_2$ complex in which bridging H-atom has the bond-length of 1.192 Å with an atomic charge of +0.568 |e|. Thus, the proton ($H^+$) is equally shared between two $H_2O$ molecules, which lies midway between them at equilibrium. This structure is in accordance with the previous reports [29, 30, 35-38], including that of Hou *et al.* [8]. Yu and Bowman [45] have shown that the structure of Zundel $O_2H_5^+$ can be used to predict the proton stretching and bending frequencies of larger protonated water clusters such as $O_4H_9^+$, $O_5H_{11}^+$, etc. In the case of $O_3H_7^+$, we obtain two low lying isomers, Eigen-1 ($C_1$) and Eigen-2 ($C_s$) with the energy difference of merely 0.1 kcal/mol, as listed in Table 1. As shown in Fig. 1, both isomers consist of an Eigen moiety, which is linked to two $H_2O$ moieties via H-bonds of 1.455-1.461 Å in Eigen-1 and 1.456 Å in Eigen-2. Thus, the structure can be thought of as an $OH_3^+$-$2H_2O$ complex in which bridging $OH_3$ moiety possesses a net charge of +0.836 |e|.

For $O_4H_9^+$, which is obtained by replacing all H-atoms of $OH_3$ by $OH_3$ itself, there exist four isomers as displayed in Fig. 2. From Table 1, one can note that the lowest energy corresponds to Eigen structure, which contains $OH_3$ moiety with a net charge of +0.825 |e| connected to three $H_2O$ moieties via H-bonds. Two Zundel type isomers, Zundel-1 and Zundel-2 being close in energy and having two H-bonds of 1.548 Å, are 3.8-3.9 kcal/mol higher in energy. Furthermore, ring isomer having three H-bonds is 7.9 kcal/mol higher in energy than the Eigen structure. Kulig and Agmon [46] suggested that both Zundel and Eigen isomers contribute to the infrared spectrum of the gas-phase $O_4H_9^+$ cluster. If we further increase the length of this series by replacing one more H-atom by $OH_3$ moiety, we get five isomers of $O_5H_{11}^+$, two ring structures, two branch structures and one chain structure as displayed in Fig. 3. Ring-1 structure, which has the lowest energy, possesses an $OH_3$ Eigen moiety with +0.820 |e| and five H-bonds. On the contrary, ring-2 having relative energy of 3.6 kcal/mol, contains an $O_2H_5$ Zundel moiety with four H-bonds. Branch-1 and branch-2, both having Eigen moiety and four H-bonds, are only 0.6 and 0.7 kcal/mol above in energy, respectively. The chain structure, which is 4.2 kcal/mol higher in energy, possesses an $O_3H_7$ type moiety with four H-bonds. These structures are in accordance with those reported earlier by Christie and Jordon [47].

### 3.2. H-bond interactions and stabilization

Above discussions suggest that except $O_2H_5^+$, $O_xH_{2x+1}^+$ cations prefer Eigen type structures as expected [35-38] having a positive charged $OH_3^+$ moiety, which is stabilized by a number of H-bonds. The experimental observations already confirmed [48, 49] that hydronium ion can form H-bonds to a number of water molecules in clusters even in the gas-phase. The characteristics of these H-bonds can be described by the quantum theory of atoms in molecules (QTAIM) analyses. QTAIM explains the bonding between two atoms by the existence of a bond critical point (BCP) and topological parameters such as charge density ($\rho$), its Laplacian ($\nabla^2\rho$), potential energy density ($V$), kinetic energy density ($G$) and total electronic energy density ($H$) at BCP. For $x > 2$, $O_xH_{2x+1}^+$ cations appear as H-bonded complexes. Molecular graph generated for $O_xH_{2x+1}^+$ ($x > 2$) by QTAIM method are displayed in Fig. 4. For simplicity of discussion, we have considered only the lowest energy structures.

According to Rozas *et al.* [50], $\nabla^2\rho > 0$ and $H < 0$ for medium H-bond of partially covalent nature but $\nabla^2\rho > 0$ and $H > 0$ for weak H-bond of electrostatic character. The topological parameters at BCP associated with H-bonds in $O_xH_{2x+1}^+$ species are also displayed in Fig. 4. It is evident that both $O_3H_7^+$ and $O_4H_9^+$ possess two and three partially covalent H-bonds, respectively. On the contrary, $O_5H_{11}^+$ possesses three partially covalent H-bonds and two electrostatic H-bonds. According to Espinosa *et al.* [51], the strength of H-bonds ($\Delta E$) can be quantified as, $\Delta E = -\frac{1}{2} V$. The $\Delta E$ values are 29.3 kcal/mol and 20.6 kcal/mol in $O_3H_7^+$ and $O_4H_9^+$, respectively, for each H-bond. Thus, these H-bonds possess medium strength as suggested by their topological parameters. Furthermore, $O_5H_{11}^+$ is stabilized by three medium H-bonds of 21.8, 21.6 and 19.2 kcal/mol as well as two weak H-bonds of 5.3 and 5.2 kcal/mol in accordance with their topological parameters (Fig. 4) and bond-distance (Fig. 3).

The stabilization of $O_5H_{11}^+$ species can be further verified by analyzing their dissociation into various fragments. We have considered two prominent dissociation pathways, such as $O_xH_{2x+1}^+ \rightarrow xH_2O + H^+$ (deprotonation) and $O_xH_{2x+1}^+ \rightarrow (x\text{-}1)H_2O + OH_3^+$ (dehydration). The corresponding dissociation energies $\Delta E_{\text{depro}}$ and $\Delta E_{\text{dehyd}}$ are listed in Table 1. One can see that the deprotonation of $O_xH_{2x+1}^+$ cations is quite difficult due to high $\Delta E_{\text{depro}}$ value, which goes on increasing with the increase in $x$. The $\Delta E_{\text{dehyd}}$ values are very important in the sense that $O_xH_{2x+1}^+$ cations can be realized as $OH_3^+\cdots(x\text{-}1)H_2O$ complexes. From Table 1, it is evident that $O_xH_{2x+1}^+$ cations are energetically stable against the dehydration, i.e., the loss of water molecules as all $\Delta E_{\text{dehyd}}$ values are positive. For instance, the $\Delta E_{\text{dehyd}}$ value of $O_2H_5^+$, i.e., 35.5 kcal/mol is in accordance with the previous estimate at CCSD(T)/6-311++G(3df, 3pd) level [8]. Furthermore, like $\Delta E_{\text{depro}}$, $\Delta E_{\text{dehyd}}$ values and hence, the stability of $O_xH_{2x+1}^+$ cations increases with the increase in $x$, which is consistent with their H-bond strengths described above.

### 3.3. Superalkali properties

In order to explore the superalkali nature of $O_xH_{2x+1}^+$ cations, we have calculated their vertical electron affinity ($EA_v$) and listed in Table 1. Note that the vertical electron affinities ($EA_v$) of cations are equivalent to the ionization energy (IE) of corresponding neutral species. Clearly, the $EA_v$ of $OH_3^+$ (5.16 eV) is lower than the IE of the Li atom, which is 5.39 eV [52]. Likewise, the $EA_v$ of $O_2H_5^+$ is comparable to the IE of the Cs atom (3.89 eV). Note that the Hou et al. [8] have estimated the $EA_v$ of $O_2H_5^+$ as 3.90 eV using OVGF/6-311++G(3df, 3pd) level. Furthermore, the $EA_v$ becomes 3.22–3.23 eV for $O_3H_7^+$ and 2.78–3.04 eV for $O_4H_9^+$. Moreover, the $EA_v$ of $O_5H_{11}^+$ is further reduced to 2.50–2.79 eV with that of its lowest energy structure as 2.67 eV. Therefore, $O_xH_{2x+1}^+$ species form a new series of non-metallic superalkali cations.

In Fig. 5, we have shown the variation of the $EA_v$ of (the lowest energy) $O_xH_{2x+1}^+$ species as a function of the number of O atoms ($x$). One can see that the difference of vertical electron affinities ($\Delta EA_v$) between two consecutive species of $O_xH_{2x+1}^+$ series goes on decreasing with the increase in $x$. For instance, the $\Delta EA_v$ between $OH_3^+$ and $O_2H_5^+$ is 1.29 eV which is reduced to 0.65 eV for that between $O_3H_7^+$ and $O_4H_9^+$. Evidently, the $EA_v$ of $O_xH_{2x+1}^+$ cations decreases rapidly with an increase in $x$. We can expect even lower $EA_v$ value with smaller $\Delta EA_v$ for $x > 5$. Therefore, it seems interesting to analyze the extreme (lowest) value of $EA_v$ of $O_xH_{2x+1}^+$ cations.

Considering the fact that the $O_xH_{2x+1}^+$ cations can be supposed to be $OH_3^+\cdots(x\text{-}1)H_2O$ ionic complexes, we have analyzed the effect of water solvent on the $EA_v$ of $OH_3^+$ using polarisable continuum model (PCM). According to PCM, the variation of the free energy ($G$) when going from vacuum to solution is composed of the work required to build a cavity in the solvent (cavitation energy) together with the electrostatic and non-electrostatic works such as dispersion and repulsion [53]. The equilibrium structure of $OH_3^+$ trapped within cavity of water is displayed in Fig. 6. We have calculated free energy of solvation ($\Delta G_{sol}$) as the difference of the free energy of $OH_3^+$ in water and that in vacuum or gas phase. $\Delta G_{sol}$ gives the free energy required to transfer a solute molecule from solvent to vacuum or gas phase. The $\Delta G_{sol}$ for $OH_3^+$ in water solvent is found to be -67.7 kcal/mol. This free energy barrier confirms the solubility of $OH_3^+$ cation in water solvent. The $EA_v$ of solvated $OH_3^+$ molecule is calculated to be 1.85 eV, which provides a lower limit of $EA_v$ value of $O_xH_{2x+1}^+$ cations at least approximately.

## 4. Conclusions

We have systematically studied a series of $O_xH_{2x+1}^+$ cations ($x$ = 1-5) and explored their structural features, interactions and superalkali behavior using MP2/6-311++G(d,p) level. We have noticed that $O_xH_{2x+1}^+$ cations energetically favour Eigen type moieties for all the structures studied except $O_2H_5^+$, which is Zundel structure. QTAIM analyses revealed that these cations are stabilized by a number of partially covalent and electrostatic H-bonds, and can be expressed in the form of $OH_3^+\cdots(x\text{-}1)H_2O$ complexes. The interaction energies of these medium and weak H-bonds lie in the range 19.2-29.3 kcal/mol and 5.2-5.3 kcal/mol, respectively. These cations are found to be stable against deprotonation [$O_xH_{2x+1}^+ \rightarrow xH_2O + H^+$] with the energy 172.7-269.1 kcal/mol as well as dehydration [$O_xH_{2x+1}^+ \rightarrow (x\text{-}1)H_2O + OH_3^+$] with the energy 35.5-96.4 kcal/mol. Interestingly, $O_xH_{2x+1}^+$ form a new series of non-metallic superalkali cations, whose vertical electron affinity ($EA_v$) varies between 5.16 eV to 2.67 eV as $x$ varies from 1 to 5. This series may be continued to include a new member with even lower $EA_v$ values. The limiting value of the $EA_v$ of $O_xH_{2x+1}^+$ cations has been approximated to 1.85 eV by considering the effect of water solvent on $OH_3^+$ implicitly using polarizable continuum model. We believe that the findings of this study will provide new insights into the nature of protonated water clusters and motivate their applications in the design of novel systems with various interesting properties.

## Acknowledgement

Dr. A. K. Srivastava acknowledges Prof. S. N. Tiwari, Department of Physics, DDU Gorakhpur University, for helpful discussions and University Grants Commission (UGC), New Delhi, India for approving Start Up project.

Table 1. The relative energy ($\Delta E_{rel}$), deprotonation energy ($\Delta E_{depro}$), dehydration energy ($\Delta E_{dehyd}$) and vertical electron affinity ($EA_v$) of $O_xH_{2x+1}^+$ cations and their isomers calculated at MP2/6-311++G(d,p) level.

| $x$ | Isomer | Sym. | $\Delta E_{rel}$ (kcal/mol) | $\Delta E_{depro}$ (kcal/mol) $\rightarrow xH_2O + H^+$ | $\Delta E_{dehyd}$ (kcal/mol) $\rightarrow (x\text{-}1)H_2O + OH_3^+$ | $EA_v$ (eV) |
|---|---|---|---|---|---|---|
| 1 | - | $C_{3v}$ | - | 172.7 | - | 5.16 |
| 2 | - | $C_2$ | - | 208.2 | 35.5 | 3.87 |
| 3 | Eigen-1 | $C_1$ | 0 | 233.0 | 60.3 | 3.22 |
| | Eigen-2 | $C_s$ | 0.1 | 232.9 | 60.2 | 3.23 |
| 4 | Eigen | $C_1$ | 0 | 253.8 | 81.1 | 2.78 |
| | Zundel-1 | $C_1$ | 3.8 | 250.0 | 77.3 | 3.00 |
| | Zundel-2 | $C_2$ | 3.9 | 249.9 | 77.2 | 2.80 |
| | Ring | $C_s$ | 7.9 | 245.9 | 73.2 | 3.04 |
| 5 | Ring-1 | $C_1$ | 0 | 269.1 | 96.4 | 2.67 |
| | Branch-1 | $C_1$ | 0.6 | 268.5 | 95.8 | 2.50 |
| | Branch-2 | $C_1$ | 0.7 | 268.4 | 95.7 | 2.50 |
| | Ring-2 | $C_1$ | 3.6 | 265.5 | 92.8 | 2.79 |
| | Chain | $C_1$ | 4.2 | 264.9 | 92.2 | 2.52 |

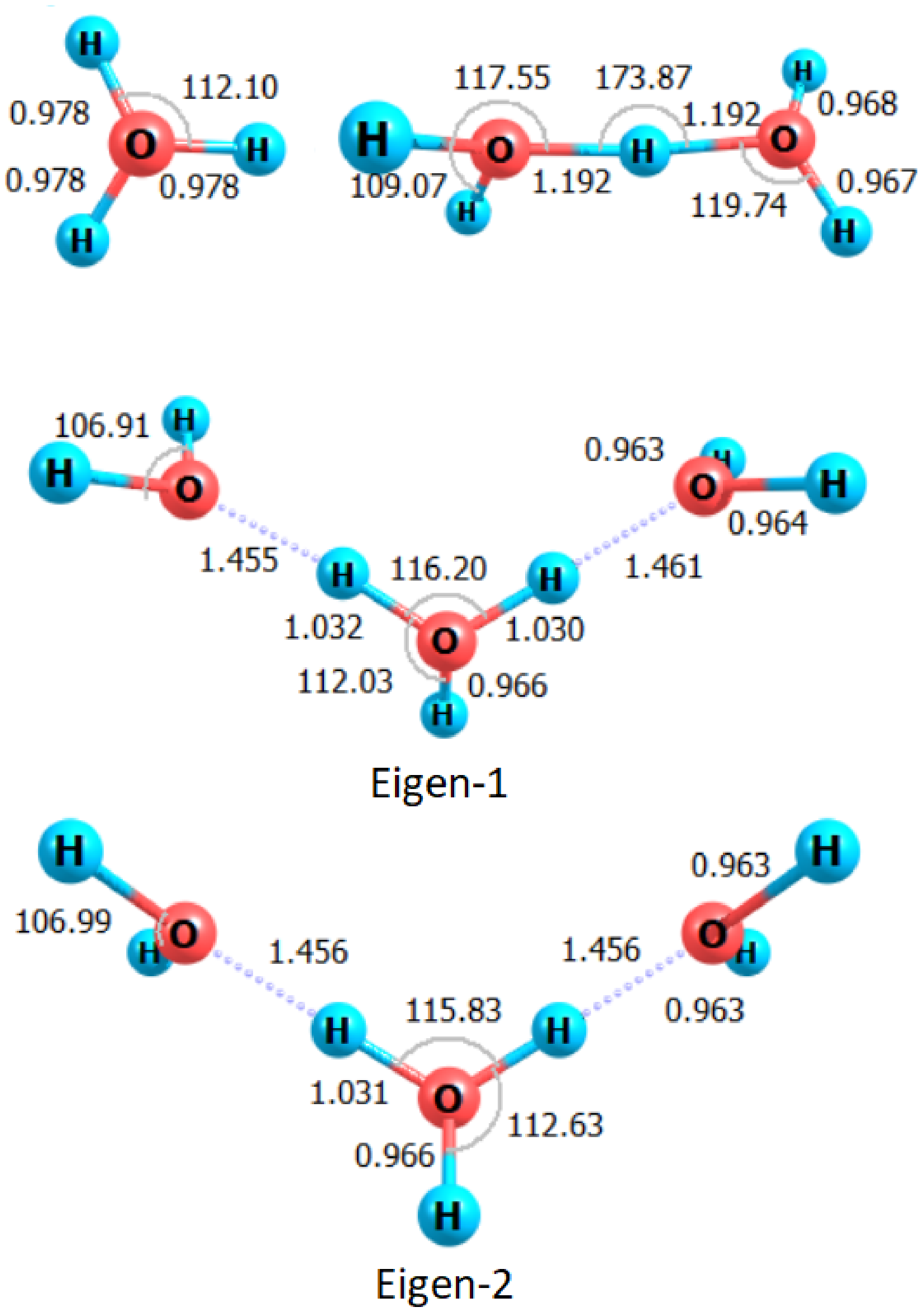

Fig. 1. Equilibrium structures of $O_xH_{2x+1}^+$ cations ($x$ = 1-3) obtained at MP2/6-311++G(d,p) level.

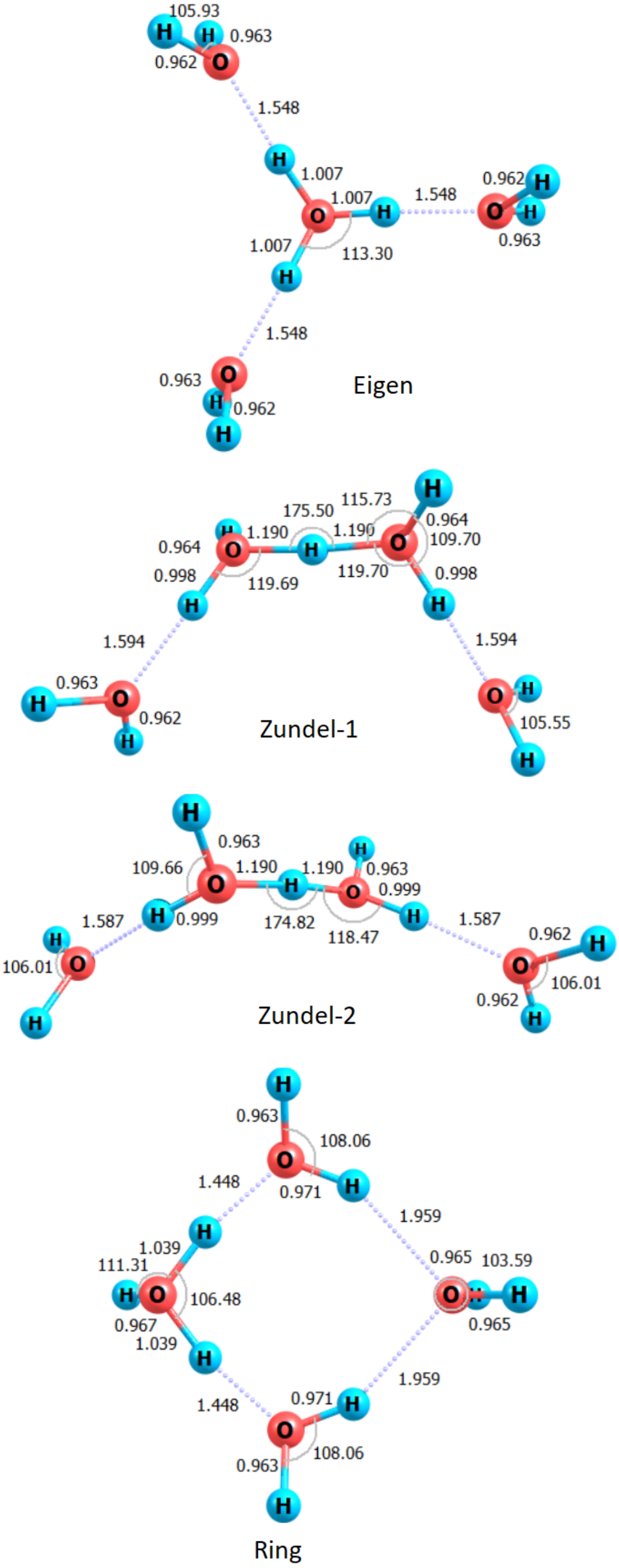

Fig. 2. Equilibrium structures of $O_4H_9^+$ cation and their isomers obtained at MP2/6-311++G(d,p) level.

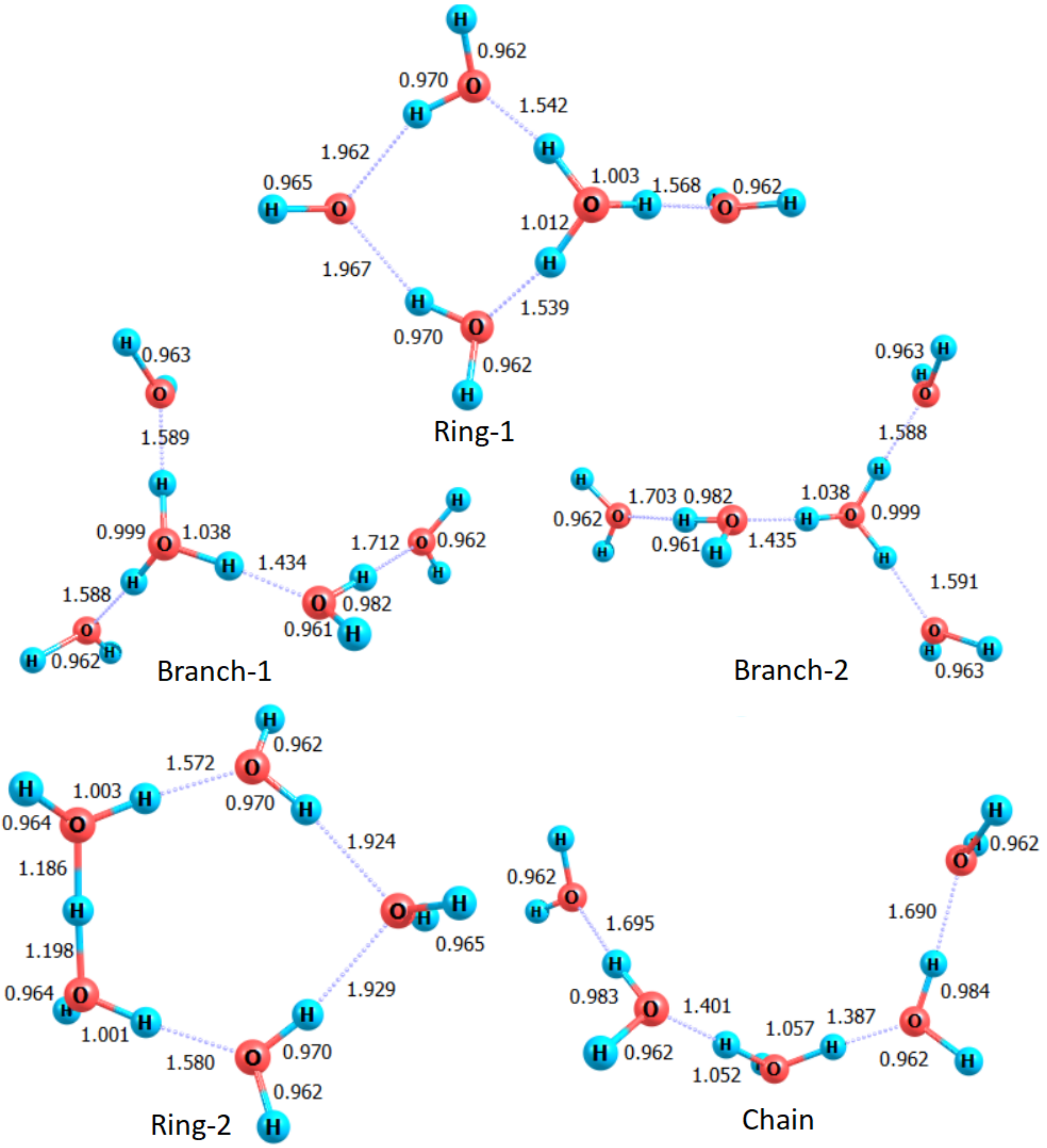

Fig. 3. Equilibrium structures of $O_5H_{11}^+$ cation and their isomers obtained at MP2/6-311++G(d,p) level.

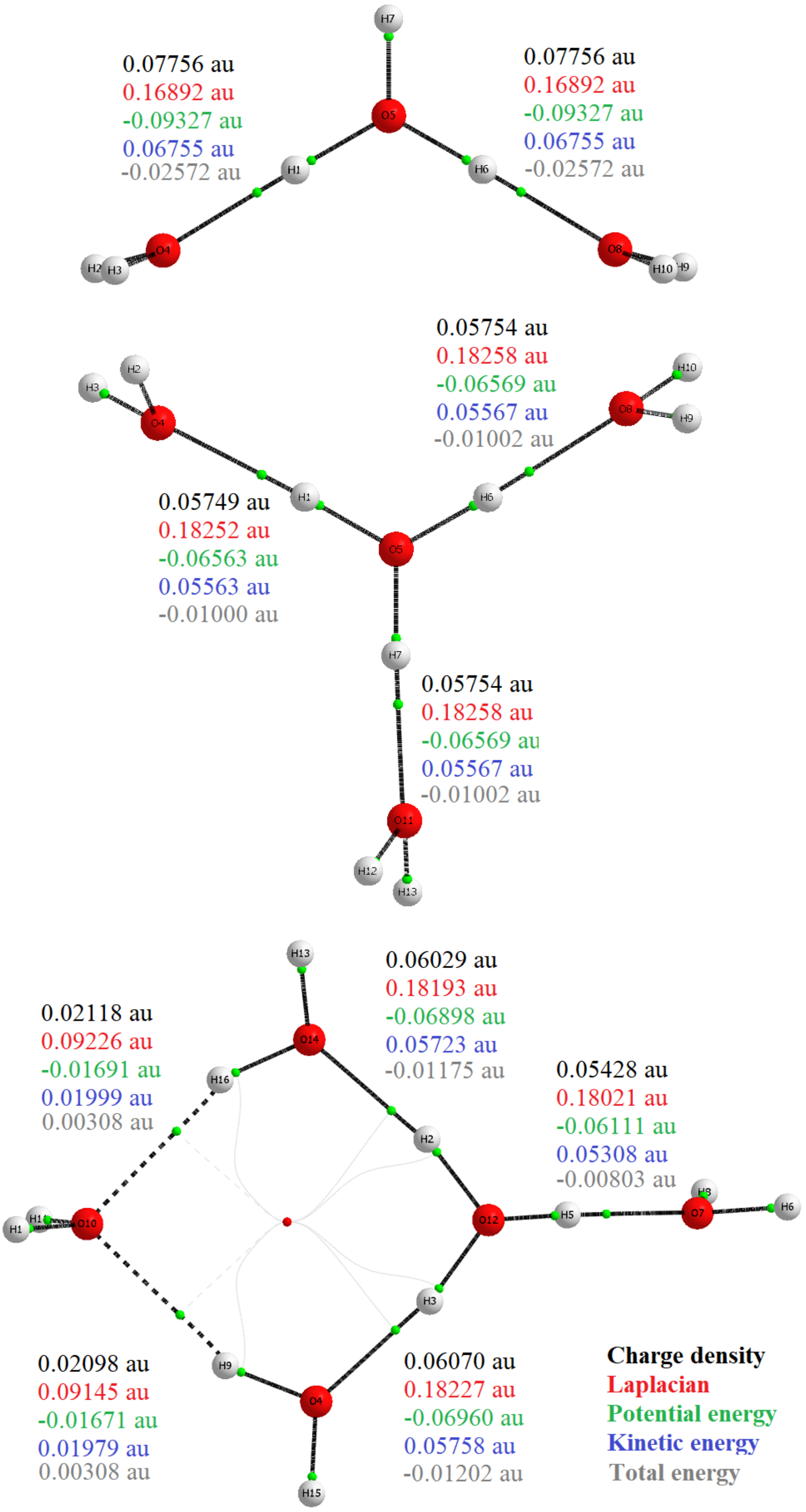

Fig. 4. Molecular graph of the lowest energy $O_3H_7^+$, $O_4H_9^+$ and $O_5H_{11}^+$ structures with topological parameters at BCP (green points) computed by QTAIM method.

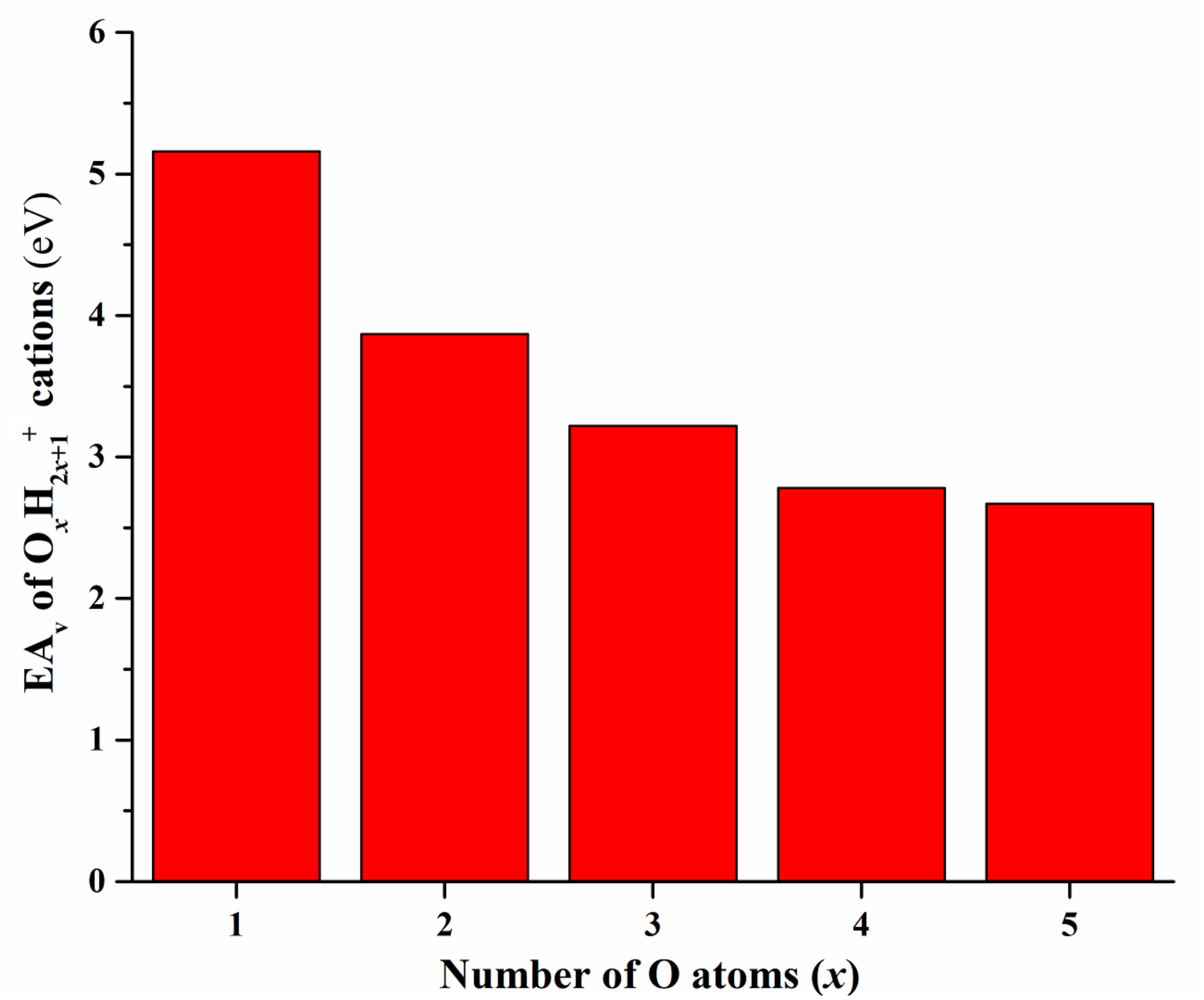

Fig. 5. Variation of vertical electron affinity ($EA_v$) of $O_xH_{2x+1}^+$ cations as a function of $x$.

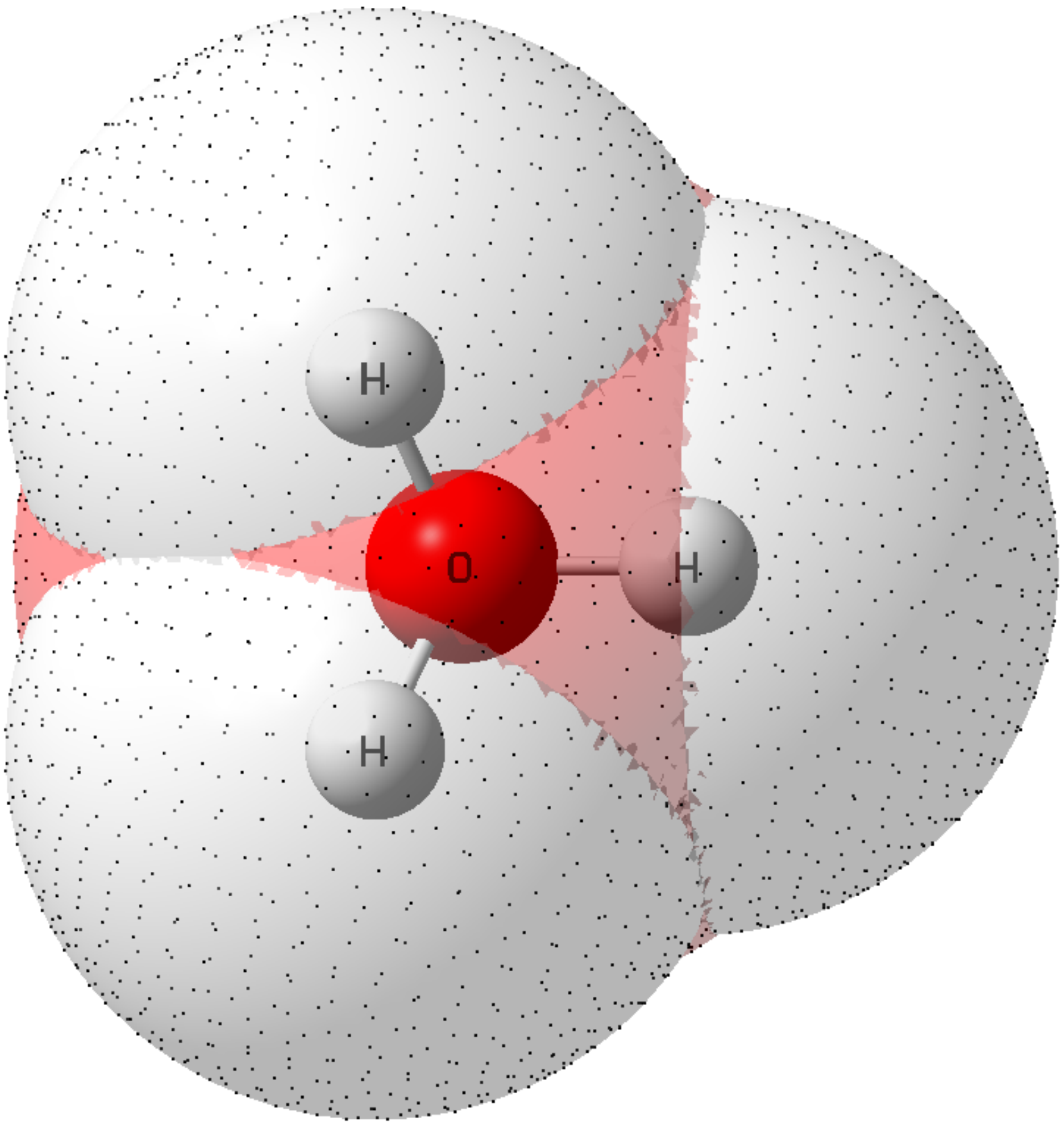

Fig. 6. $OH_3^+$ trapped within the cavity of water in polarisable continuum model.